\documentclass[pra,aps, twocolumn,floatfix]{revtex4}

\usepackage{amsfonts}\usepackage{amssymb}
\usepackage{amsmath}\usepackage{graphicx}
\usepackage{textcomp}
\usepackage{epsfig}
\usepackage{color,soul}

\begin{document}
\title{Angle-resolved polarimetry measurements of antenna-mediated fluorescence}
\author{Abbas Mohtashami, Clara I. Osorio and A. Femius Koenderink }
\affiliation{Center for Nanophotonics, FOM Institute AMOLF, Science Park 104, 1098 XG Amsterdam, The Netherlands}
 
\begin{abstract}
Optical phase-array antennas can be used to control not only the angular distribution but also the polarization of fluorescence from quantum emitters. The emission pattern of the resulting system  is determined by the properties of the antenna, the properties of the emitters and the strength of the antenna-emitter coupling. Here we show that Fourier polarimetry can be used to characterize these three contributions.  To this end, we measured the angle and Stokes-parameter resolved emission of  bullseye plasmon antennas as well as spiral antennas excited by an ensemble of emitters.  We estimate the antenna-emitter coupling on basis of the degree of polarization, and determine the effect of anisotropy in the intrinsic  emitter orientation  on polarization of the resulting emission pattern.  Our results not only provide new insights in the behavior of bullseye and spiral antennas, but also demonstrate the potential of Fourier polarimetry when characterizing antenna mediated fluorescence.
\end{abstract}
 
\maketitle

\section{Introduction} 

Engineering the photonic environment of  single emitters allows controlling their emission rate,  angular distribution and polarization. According to Fermi's golden rule, the emission rate depends on the local density of states (LDOS), which can be manipulated with narrowband devices, such as microcavities \cite{Englund_PRL05,Noda_NP07,Reithmaier_N04}, or with broadband structures, which include plasmonic structures \cite{Yuan_ACIE13,Kinkhabwala_NP09,Wenger_JPCC07}. On the other hand, the angular distribution of emission from single emitters can be controlled by properly-designed nano-structures that, after being excited by a point-like emitter, act as a phased arrangement of coherent secondary sources \cite{AgioAlu_book}. This approach, which is analogous to    phased-array engineering in radio frequency,  has been demonstrated  for single molecules coupled to scanning probe tips~\cite{Gersen_PRL00}, plasmonic lattices~\cite{Vecchi_PRL09,Rodriguez_PRL12,Lozano_LSA13},  patch antennas~\cite{Belacel_NL13,Mohtashami_ACSP14}, array antennas \cite{Curto_Sc10} and plasmonic bullseye antennas~\cite{Lezec_SC02,Wenger_OE05,Wenger_OE08,Wenger_IJMS10,Mahboub_OE10,Aouani_NL11,Aouani_NL11b}.  Moreover, with the advent of metasurfaces and metamaterials, it has become evident that carefully shaped structures can simultaneously, although not independently, control the electric and magnetic components of the near-field around an emitter~\cite{Sersic_PRL12,Schaferling_PRX12}.  Conversely, it has been shown that magnetoelectric scatterers can generate handed far-field emission when excited locally by an electric dipole source~\cite{Bernal_NJP13,Kruk_ACSP14}. Plasmonic structures are, therefore, a promising way to simultaneously control the phase-front and polarization of fluorescent sources.

\begin{figure}[tbh!]
 \includegraphics[scale=0.22,trim=0 100 0 0,clip=true]{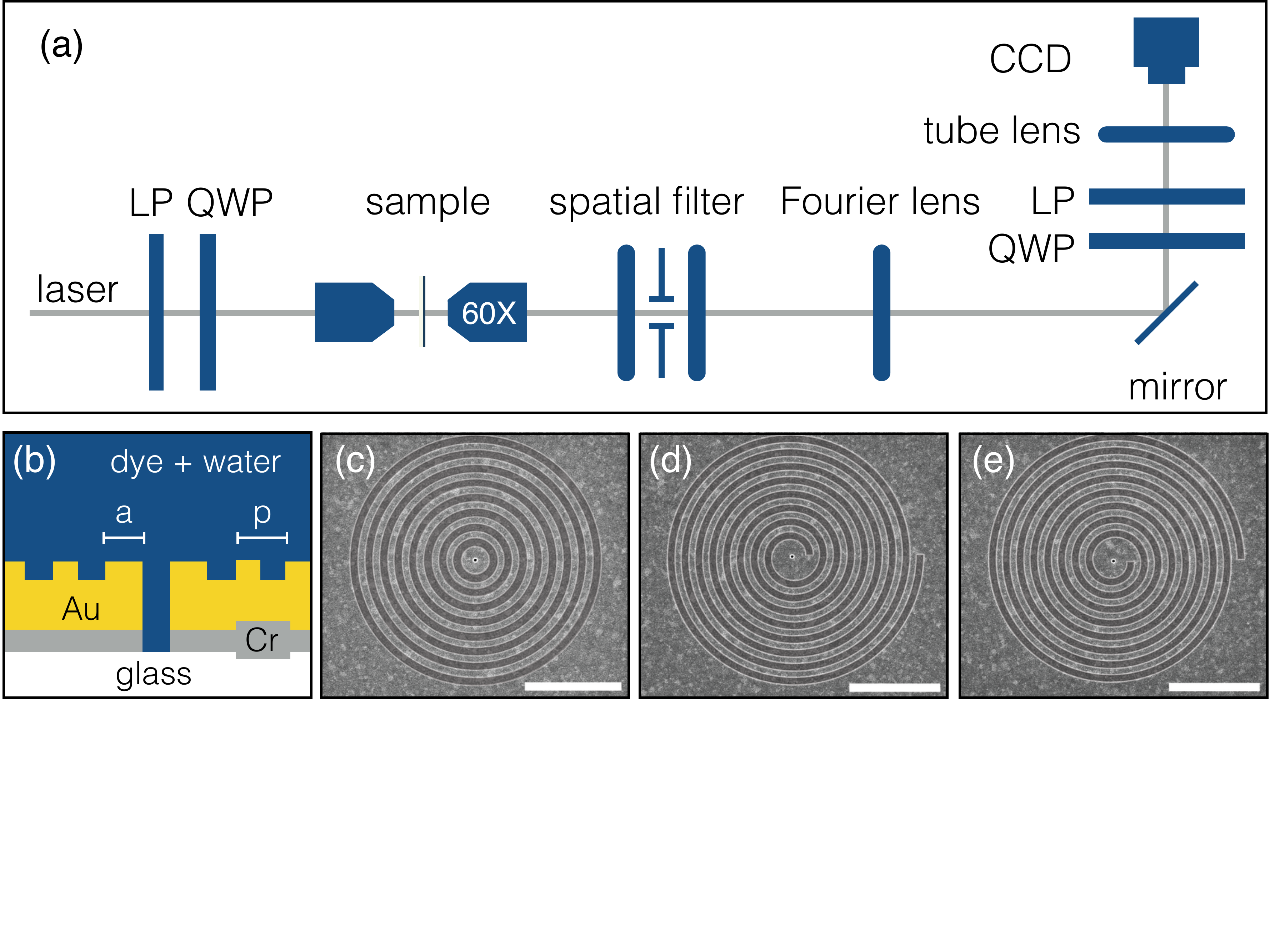}
 \caption{ (a) The experimental setup is based on  a  fluorescence Fourier microscope. The polarization of the incident light is set by a linear polarizer (LP) and a quarter-wave plate (QWP), after which the light is focused into the sample. The resulting fluorescence is then collected by a 60$\times$ microscope objective ($NA=0.7$), a spatial filter allows selecting particular regions in the sample plane. The Fourier and tube lenses are placed such that the back focal plane of the objective is imaged into the camera sensor. Finally, as polarimeter we use a quarter waveplate and a linear polarizer. (b) Crosscut of the sample, rotated $90^{\circ}$  with respect to (a). The dye molecules diffuse freely on top of each sample in a water solution contained by a flow cell. Scanning electron micrographs (SEM) show the fabricated (c) bullseye, (d) anti-clock wise spiral and (e) clock wise spiral plasmonic nano-antennas. In this SEM images, a mask was used to increase the contrast. The scale bars correspond to $4$ $\mu$m.} \label{fig_setup}
\end{figure}

An important consideration when using plasmonic structures for polarization control of fluorescence is that, in many  experiments, fluorescence is only partially polarized. If an experiment uses a single and rigidly oriented dipole emitter emission will be fully polarized \cite{Curto_Sc10,Kruk_ACSP14}. However, if the fluorescence is due to a randomly oriented ensemble of molecules, or if the molecular source is free to rapidly re-orient from  emission event to emission event, the emission will be at best partially polarized ~\cite{Lakowicz_book}. A relevant figure of merit for a plasmonic structure is hence in how far it is able to imprint polarization on an intrinsically poorly polarized source ensemble.  This requires on the one hand  a plasmon antenna with a strongly polarization-selective resonance and, on the other hand, a strong structure-emitter coupling, so that the polarized structure-mediated emission exceeds the direct emission of light into the far-field that occurs for poorly coupled emitters.   Here we show that by measuring the angle-resolved full polarization state of the emission it is possible to  separate direct emission from emission mediated by a photonic structure and,  therefore, to estimate the structure-emitter coupling. We use a k-space polarimeter that combines a Fourier microscope with a polarimeter \cite{Fallet_MEMS11,Arteaga_OE14,Kruk_ACSP14,Osorio_SR15}, to retrieve the Stokes parameters ($S_0$, $S_1$, $S_2$ and $S_3$) of  fluorescence emitted in the vicinity of bullseye antennas and spirals \cite{Mahboub_OE10,Aouani_NL11,Aouani_NL11b,Gorodetski_PRL13}.  The Stokes parameters allow us to calculate the ratio of polarized to unpolarized light, to separate those contributions to the total intensity and to calculate the electric field components that describe the polarized part \cite{BornWolf_book,Bass_book,Berry_AO77}.  


\section{Experimental methods}
\subsection{Setup}
Figure~\ref{fig_setup}(a) shows our experimental setup  composed of  a fluorescence microscope equipped with an optional Fourier transforming lens ($f_\mathrm{Fourier}=200$\,mm) and a rotating-plate polarimeter. As excitation light source, we use a broadband super-continuum laser (Fianium) filtered by an acousto-optical tunable filter (AOTF) and a bandpass filter ($680\pm 10$\,nm). The polarization of the excitation is set to horizontal or right-handed circular using a linear polarizer and a quarter-wave plate. A $10\times$ objective ($NA=0.25$, Nikon Plan) focuses the light on the  sample, where it serves to pump near-infrared fluorophores embedded in the sample. A  $60\times$ objective ($NA=0.7$, Nikon CFI Plan Fluor) collects  the fluorescence and directs it to a spatial filter.  The spatial filter is composed of a 1:1 telescope (two lenses of $f_\mathrm{telescope}=50$\,mm) and a  $300~\mu$m pinhole, which selects an area of about 20\,$\mu$m across in the sample plane. The excitation light and the fluorescence are separated by a  690\,nm long-pass filter. A section of the broad fluorescence spectrum is selected by a bandpass filter centered at $750$\,nm with a full-width at half-maximum of $40$\,nm. Finally, a polarimeter consisting of a quarter-wave plate and a linear polarizer is placed before the $f=200$\,mm tube lens and the silicon CCD camera (Photometrics CoolSnap EZ). The exposure time of the camera varied between $40$\,s and $60$\,s depending on the structure, with a constant pump power of $10$$~\mu$W.

\subsection{Samples}
We fabricated cylindrical and spiral bullseye antenna   samples from  a  200\,nm thick layer of gold evaporated on top of a glass coverslip covered by a 5\,nm thick  chromium adhesion layer. Using a focused ion beam (FIB), we
milled a 200\,nm diameter hole through the gold layer and $\sim$50\,nm deep trenches concentric to the hole. The grooves have a 50\% duty cycle.  We study bullseye antennas as well as  anti-clock and clock wise  Archimedean spiral antennas (Fig.~\ref{fig_setup}(c), (d) and (e), respectively). For each structure, the distance between the center and the first groove is $a=330$\,nm and  between consecutive grooves is $p=600$\,nm. For reference, we also fabricated single holes with no structures around.  On top of the structures,  we mounted a  flow-cell containing {\it Alexa Fluor 700} dye molecules dissolved in water at a concentration of 10\,$\mu$M, which have an emission peaked at 723\,nm with a full-width at half-maximum of about 50\,nm.  On basis of scattering measurements \cite{Osorio_SR15}, we anticipate the bullseye antennas to imprint directionality in form of a narrow donut beam on fluorescence emitted by molecules in the central aperture \cite{Lezec_SC02,Aouani_NL11,Langguth_ACSN13}.

\subsection{Polarimetry}
In order to determine the polarization state of  light emitted by dye molecules in vicinity of the nanoantennas, we retrieved the angle-resolved Stokes parameters $S_0$, $S_1$, $S_2$ and $S_3$. This procedure requires measuring Fourier images of the fluorescence intensity $I_n$ transmitted by the polarimeter when performing as a linear polarizer (horizontal, vertical, $45^{\circ}$ and $135^{\circ}$) and as a circular polarizer (right and left handed). The Stokes parameter $S_0$, equal to the total intensity of the fluorescence,  is given by the sum of any pair of orthogonally polarized intensities (e.g. $S_0=I_H+I_V$). On the other hand, the difference between each pair of orthogonally polarized intensities determine the other Stokes parameters ($S_1=I_H-I_V$, $S_2=I_{45}-I_{135}$ and $S_3=I_{RHC}-I_{LHC}$)~ \cite{BornWolf_book}. When normalized to the total intensity, these last three Stokes parameters take values between $-1$ to $1$. The two extreme values correspond to emission fully polarized in one of the two orthogonal polarizations  used to define the parameter, while $0$ corresponds to the case  where both polarizations contribute equally to the total emission.

\begin{figure*}[th!]
\begin{center}
 \includegraphics[width=120mm]{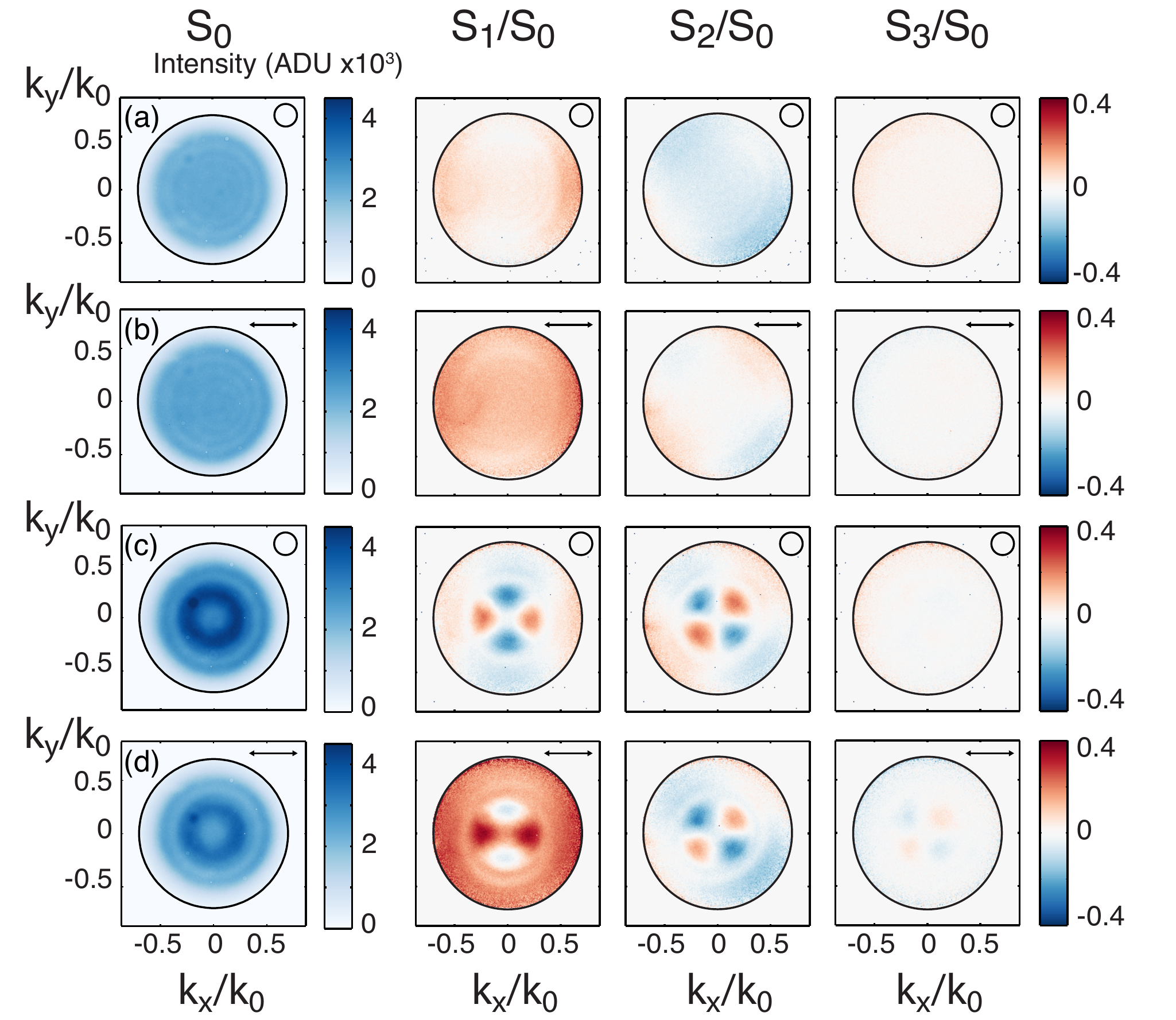}
 \caption{k-space distribution of retrieved Stokes parameters $S_0$ and  $S_1/S_0$, $S_2/S_0$ and $S_3/S_0$, for the emission of a single hole aperture (a)-(b) and a bullseye (c)-(d) excited with circularly (a)-(c) and linearly polarized light (b)-(d), symbolized by circle or an arrow in each image. In all cases, the detection wavelength is $750$ nm. Images are clipped at the NA of our objective (black circle at  $0.7$).} \label{fig_Stokes}
\end{center}
\end{figure*}

\subsection{Calibration of dye properties}
Prior to measuring on plasmonic structures, we measured  the Stokes parameters from light emitted by the dye far from any structure. Independently of the incident polarization, the normalized Stokes parameters $S_1/S_0$, $S_2/S_0$, and $S_3/S_0$ of the fluorescence show constant values throughout k-space. For incident circularly polarized light, the angular average of each of these parameters is close to zero ($\langle S_1/S_0 \rangle  =1.3\times 10^{-2}$, $\langle S_2/S_0\rangle =0.1\times 10^{-2}$, and $\langle S_3/S_0\rangle =0.1\times 10^{-2}$) indicating unpolarized emission,  as  expected for  an isotropic ensemble of emitters. On the other hand, the molecules clearly show fluorescence anisotropy~\cite{Lakowicz_book} when excited by linearly polarized light. In this case, the emission is significantly biased towards the incident horizontal polarization, with an angular average of the first Stokes parameter $\langle S_1/S_0\rangle=0.23$ (while $\langle S_2/S_0\rangle=1.2\times 10^{-4}$ and $\langle S_3/S_0\rangle=3.1\times 10^{-2}$).  For a random ensemble of linear dipolar emitters (with emission dipole along the absorption dipole) a complete absence of rotational diffusion would imply $\langle S_1/S_0\rangle\leq 0.33$~\cite{Lakowicz_book}. Thus, the significant measured bias $\langle S_1/S_0\rangle= 0.23$ implies a  rotational diffusion time that is comparable to the fluorescence lifetime ($1$ ns) of Alexa Fluor 700 molecules.  The dye ensemble retains a degree of polarization upon linear, but not circular, polarization, as a consequence of the fact that linear absorption dipoles have no memory of the \textit{phase difference} between driving vector components.

A key point of our paper is that the fluorescence anisotropy must be accounted for when interpreting plasmonically modified radiation patterns. By choosing a dye system with a short   fluorescence lifetime, in our experiments we can essentially probe our plasmonic system in two distinct ways. On one hand we can  drive the antennas with a completely unpolarized ensemble of emitters by applying circular input polarization. On the other hand, we can drive the system with a preferentially linearly polarized ensemble of emitters by pumping with a linear input polarization. As we will show, this results in completely different polarization emission patterns.  We note that a third class of experiments with fully oriented driving  can only be achieved with single molecules or physically aligned dipoles.




\section{Bullseye measurements}
We examine the angle-resolved polarization state of the light generated by molecules in a single nanoaperture and in a bullseye structure.  Figure~\ref{fig_Stokes} shows the retrieved Stokes parameters for fluorescence  emitted when the structures are illuminated with circularly,  (a)-(c), and linearly polarized light,  (b)-(d).  Single hole apertures, in Fig.~\ref{fig_Stokes} (a) and (b), show angularly isotropic emission pattern $S_0$ independent of the incident polarization~\cite{Aouani_NL11,Langguth_ACSN13}. The other Stokes parameters and therefore the polarization also show little angular structure.   There is a small polarization bias, see for instance $S_1$ and $S_2$ in Fig.~\ref{fig_Stokes} (a), due the polarizing effects of optical elements placed before the polarimeter, such as the high-NA microscope objective and a silver mirror. Horizontally polarized excitation, however, increases the emission of horizontally polarized photons as shown by the increase of the angular average of $S_1$ from  $\langle S_1/S_0 \rangle =3.0\times 10^{-2}$ for circularly polarized excitation to  $\langle S_1/S_0 \rangle =0.15$ for linearly polarized excitation.

In contrast to single holes, bullseye structures show very directional emission patterns $S_0$, which result from far-field interference of direct emission, and emission that is first funneled into plasmons and subsequently outcouples at the antenna grooves~\cite{Aouani_NL11,Aouani_NL11b}. Moreover, the clear structure in the Stokes parameters $S_1$ and $S_2$ in Fig.~\ref{fig_Stokes} (c) and (d) indicates three important features of the polarization state of the emitted light. First, even when using circularly polarized excitation to pump a random isotropic ensemble of molecules, there are angular regions were $S_1/S_0$ and $S_2/S_0$  are nonzero. Therefore the emitted light is at least partially polarized due to interaction with the plasmon antenna. Second, the clover-leaf pattern with alternating signs in $S_1$ and $S_2$ suggest that the imparted polarization is largely radial as expected due to the role of plasmons that, after being excited by the molecules,  propagate radially from the central hole to the grooves, and subsequently scatter out in a narrow donut by diffraction at the grooves. Since plasmons are TM-waves~\cite{Novotny_book,Maier_book}, one expects the outcoupled donut to have a well defined radial  polarization. Finally, for all angles of emission $S_3/S_0\approx 0$, meaning that the measured light shows no circularly polarized component. This stands to reason since both contributions to the emission, i.e., direct emission by the molecules and plasmon scattering from the grooves, are either linearly polarized or unpolarized. In contrast, if one performs scattering experiments on  bullseyes~\cite{Osorio_SR15}, linearly polarized input light can be scattered as circularly polarized light at non-normal angles. This is possible  since the scattering includes a coherent superposition of   the diffracted pump beam and light scattered by the bullseye groove~\cite{Rodriguez_Sc13}.

\begin{figure}[t!]
 \begin{center}
  \includegraphics[width=85mm]{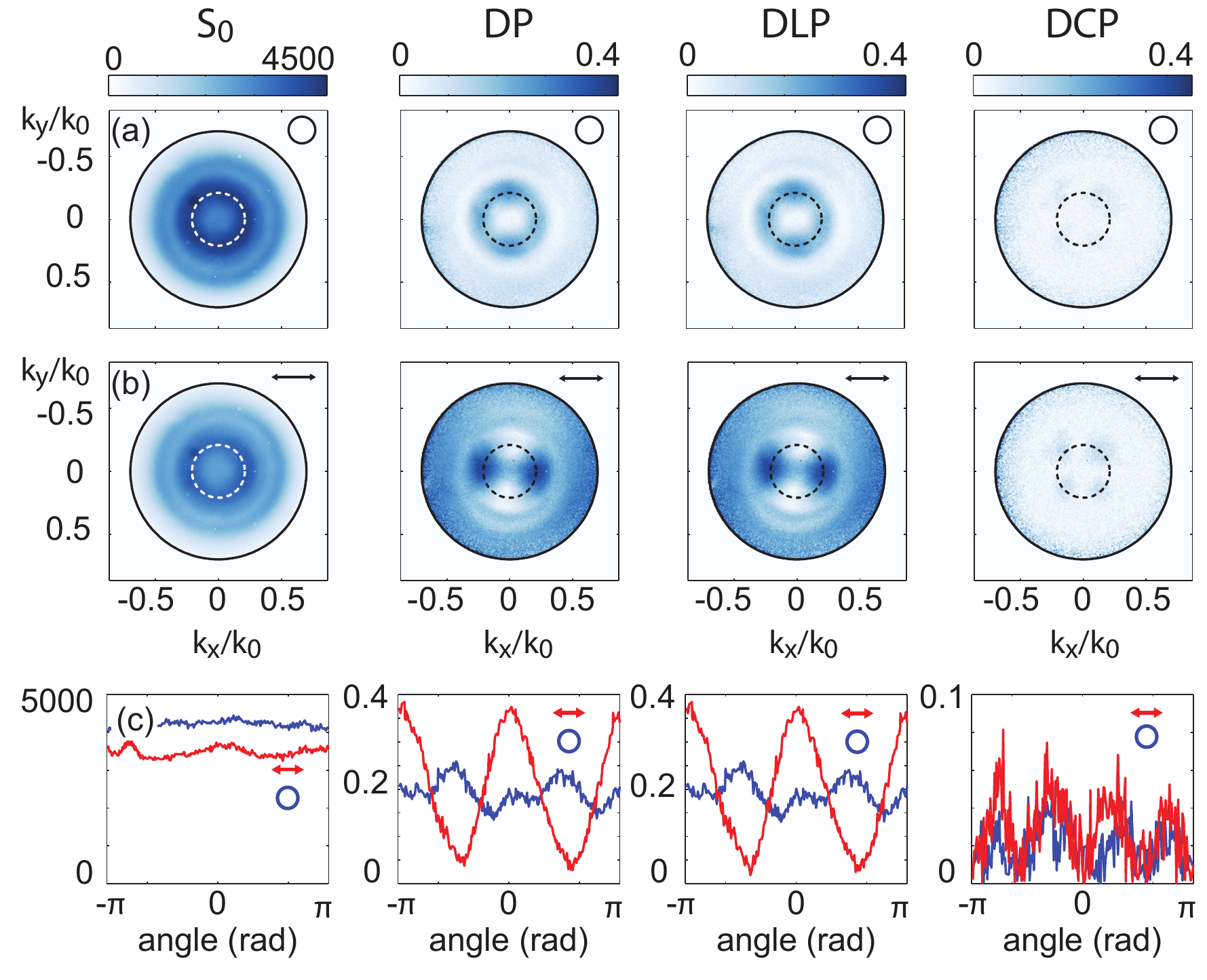}
  \caption{Emission pattern ($S_0$), total degree of polarization ($DP$), degree of linear polarization ($DLP$), and degree of circular polarization ($DCP$) for a bullseye structure illuminated by circularly polarized   (a) and linearly polarized light  (b). The crosscuts in (c) compare each of the quantities at  $|\mathbf{k}|/|k_0|=0.2$ (marked with a dotted circle in (a) and (b)), for circularly (blue) and linearly (red) polarized excitation. } \label{fig_DP}
  \end{center}
 \end{figure}

Next,  we convert measured Stokes parameters into the degree of polarization of the emitted light, i.e. the ratio of polarized light to total intensity, as well as the degrees of linear ($DLP$) and circular ($DCP$) polarization, according to
\begin{equation}\label{Eq_DoP}
    \begin{array}{l c l}
        DP & = & \cfrac{\sqrt{S_1^2+S_2^2+S_3^2}}{S_0}\\
        DLP & = & \cfrac{\sqrt{S_1^2+S_2^2}}{S_0}\\
        DCP & = & \cfrac{|S_3|}{S_0}.\\
    \end{array}
\end{equation}
Figure~\ref{fig_DP} shows these three quantities as well as the total emission $S_0$ for light emitted by fluorophores in the vicinity of the bullseye antenna, when excited with circularly (a) and linearly polarized light (b).    In all instances the degree of circular polarization $DCP\approx0$  is negligible, while the degree of \emph{linear} polarization $DLP$ is, within error, identical to the total degree of polarization $DP$. The figure shows that the $DP$ depends on the incident polarization, with maximum values going from  $DP_{max}=0.25$  for circular polarization excitation to $DP_{max}=0.39$   for linear polarization.  These  maximum values of $DP$ are set by the fraction of molecules actually coupled to the plasmon resonance. In the case of circularly polarized excitation, $DP_{max}$ is a direct measurement of the percentage of light emitted  into the far-field via  the antenna,  while for incident linear polarization  $DP_{max}$  sets an upper bound, given that direct emission is partially polarized.


The incident pump polarization not only influences the maximum attained degree of polarization $DP_{max}$, but also the angular distribution of the different degrees of polarization. Figure~\ref{fig_DP} (c) shows $S_0$, $DP$, $DLP$ and $DCP$ as a function of the polar angle for a fixed ratio $|\mathbf{k}|/|k_0|=0.2$ (i.e., at the donut beam opening angle of ca. $9^\circ$ in water) indicated by the dotted circles in Fig.~\ref{fig_DP} (a) and (b). While circular polarized excitation (blue) produces a ring of emission that is approximately  uniformly polarized, linearly polarized excitation (red) results in a donut beam that has strongly polarized and unpolarized lobes. These patterns result from the incoherent superposition of the homogeneous and (slightly) horizontally biased emission of molecules not coupled to the antenna,  and the radially polarized donut beam generated by the plasmon scattering. While these two contributions have the same polarization around $|k_y|/|k_0|=0$ effectively increasing $DP$,  around $|k_x|/|k_0|=0$ the radially polarized light and the background are orthogonally polarized decreasing $DP$, thereby causing the `unpolarized' lobes.  

\begin{figure}[th!]
 \includegraphics[width=85mm]{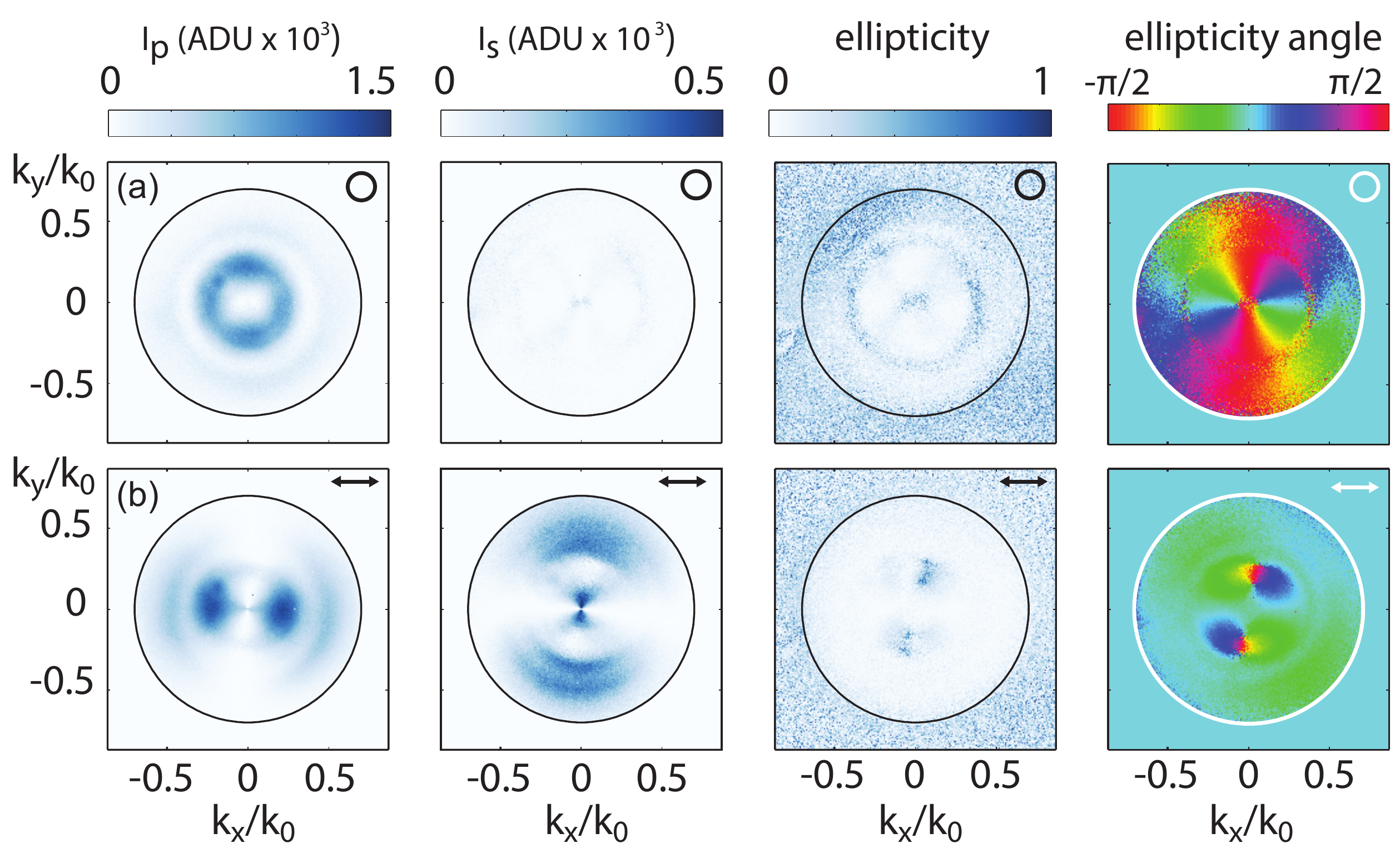}
 \caption{Intensities of $p$ and $s$ polarized field components, ellipticity and ellipticity angle of  light emitted under linearly polarized (a) and circularly polarized (b) excitation.}  \label{fig_fields}
\end{figure}

It is possible to obtain further information about the structure of the emission patterns by separating polarized from unpolarized emission, and using other figures of merit to describe the polarized part. The amplitude of the electric field components, in Cartesian coordinates $|E_x|, |E_y|$, and the phase between them, $\delta$, are given by \cite{BornWolf_book}
\begin{equation}\label{Eq_pl_ExEydelta}
\begin{array}{lll}
 |E_x|^2 & = & (S_0+S_1)/2\\
 |E_y|^2 & = & (S_0-S_1)/2\\
 \delta &=&\arg (S_2+i S_3).\\
\end{array}
\end{equation}
Transforming these cartesian backaperture field components to cylindrical coordinates  provides direct access to the $p$ and $s$ polarized components of the spherical wave emitted by the object, as evident from the ``Abbe sine condition'' transformation rules by which  objectives transform a spherical wave from an object point to a cylindrical collimated beam~\cite{Novotny_book}.  Figure~\ref{fig_fields} shows the  radial $I_p=|E_r|^2$ and tangential $I_{s}=|E_{\varphi}|^2$ intensity distributions for the \textit{polarized part} of the fluorescence generated at our bullseye antenna.  Under circularly polarized excitation the molecules' direct emission is mostly unpolarized. Therefore   Fig.~\ref{fig_fields}(a) shows the fully $p$ polarized intensity resulting from the scattering of radially propagating plasmons by the grooves~\cite{Lezec_SC02,Aouani_NL11,Mahboub_OE10}. On the other hand, the direct emission from oriented molecules appears as tangentially polarized at angles where the tangential direction coincides with the direction of the incident polarization and where the radially polarized emission from the structure does not cancel the effect, Fig.~\ref{fig_fields}(b). This behavior is confirm when retrieving other figures of merit that describe the polarized part of the emission, such as the parameters of the polarization ellipse shown in the two last columns of   Fig.~\ref{fig_fields}.  

\begin{figure}[th!]
  \includegraphics[width=85mm]{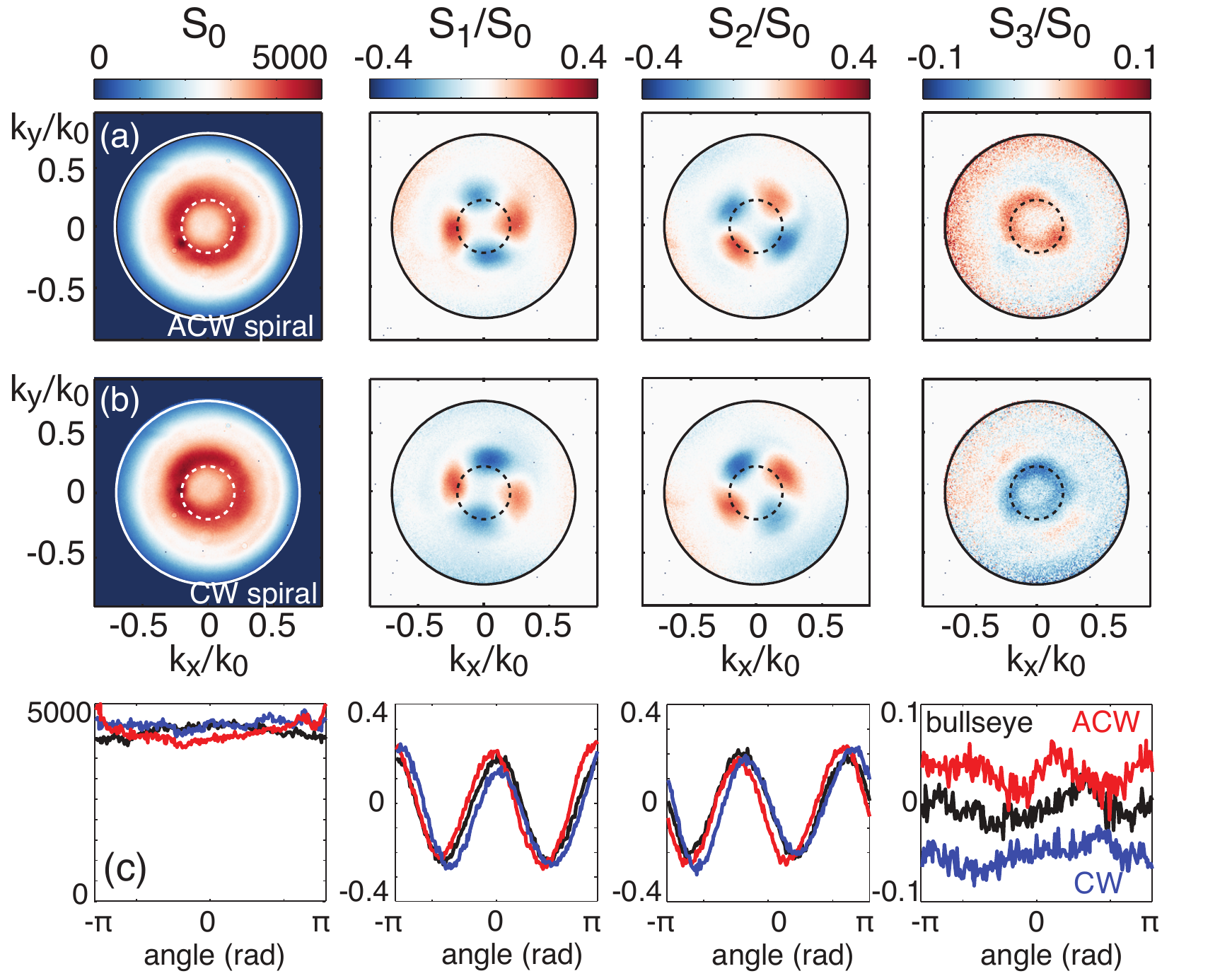}
  \caption{Angle resolved Stokes parameters $S_0$, $S_1/S_0$, $S_2/S_0$ and $S_3/S_0$ of an (a) anti-clockwise spiral  and (b) clockwise Archimedean spiral excited with circularly polarized light. For comparison, the crosscuts in (c) include the the Stokes parameters of  the bullseye under the same illumination (black lines).} \label{fig_spirals}
 \end{figure}

While bullseye mediated emission results in a linearly (radially) polarized donut beam, one would expect handed structures to impose handed polarization.  Figure \ref{fig_spirals} shows the angle-resolved Stokes parameters of light emitted in the vicinity of an anti-clockwise (a) and a clockwise (b) Archimedean spiral. As excitation we use circularly polarized pump light, which should again result in excitation of a random isotropic ensemble of dipole moments. While the total intensity distribution $S_0$ resembles closely the bullseye emission, the handedness of the structures is translated into the other Stokes parameters. The clover-leaf patterns observed in $S_1/S_0$ and $S_2/S_2$ for bullseye emission, change shape and orientation resulting in non mirror symmetric angular distributions. Crosscuts at  $|\mathbf{k}|/|k_0|=0.2$ show that the asymmetry reverses with reversal of the spiral handedness. In addition, the $S_3/S_0$ patterns show  a small amount of circularly polarized light emitted by the structures in the regions of higher total emission with a handedness given by the handedness of the structure. The crosscut coincides with the region of maximum circularly polarized emission which, however, only accounts for $4\%$ of the total emission or, equivalently,  $18\%$  of the polarized emission. Increasing this fraction separates in two challenges: on the one hand it requires improving the overall coupling strength between emitters and antenna (raising  $DP$) and on the other hand providing stronger chirality to enhance $DCP/DP$, as has been shown is the case for quantum dots coupled to split ring resonators \cite{Kruk_ACSP14}.  From a methods point of view, the data set shows the large potential of  Fourier polarimetry to determine the polarization performance of single plasmon antennas coupled to fluorophores.

\section{Conclusions}
We reported  Fourier polarimetry measurements on the emission of bullseye and spiral antennas coupled to an ensemble of \textit{Alexa Fluor 700} dye molecules dissolved in water. The measured angle-resolved Stokes parameters allows us to separate polarized from unpolarized contributions to the fluorescence, providing deeper insight in the behavior of both the emitters and the antenna, and their coupling.  In the particular dye system we chose, fluorescence decay is fast, almost on the time scale of rotational diffusion.  Thereby, we can probe both the case of a ``frozen'' anisotropic molecular ensemble, using linear excitation polarization, and a random orientational ensemble of emitters, using  circularly polarized excitation. In the latter  case, the polarized part of the emission is only due to the scattering from the bullseye, and has a radially polarized emission pattern.  In this case the degree of polarization is a measure for the emitter-antenna coupling. In contrast, under linearly polarized excitation, a very different   angular distribution of polarized and unpolarized emission is observed, showing the importance in plasmon fluorescence enhancement measurements to have a good grasp of the source orientation distribution.  Finally, we demonstrated a small amount of circularly polarized emission when using Archimedean spirals, which can likely be enhanced by increasing the chirality of the antenna and its coupling to the emitters, as shown in Ref. \cite{Kruk_ACSP14}.

\section{Acknowledgment}
This work is part of the research program of the Foundation for Fundamental Research on Matter (FOM), which is part of the Netherlands Organization for Scientific Research (NWO). This work is supported by NanoNextNL, a micro- and nanotechnology consortium of the Government of The Netherlands and 130 partners.

\bibliography{polarimetry1}

\end{document}